\documentclass[]{ametsoc}
\journal{jcli}

 \usepackage{tabularx}
\usepackage{graphicx}
\usepackage{multirow}      
\usepackage{multicol}
\usepackage{booktabs}
\usepackage[section]{placeins}
\usepackage[export]{adjustbox}
\usepackage{siunitx}
\usepackage{amssymb}

\DeclareMathOperator*{\argmin}{arg\,min}
\usepackage{soul}

\newcommand{\cpd}{$^{\circ}$C~pd }
\newcommand{\cpdend}{$^{\circ}$C~pd}
\newcommand{\kmpd}{\,\si{km^2~pd} }
\newcommand{\kmpdend}{\,\si{km^2~pd}}
\usepackage{soul}
\usepackage{color}
\usepackage{array}
\newcolumntype{R}[1]{>{\raggedleft\arraybackslash}p{#1}}
\newcolumntype{L}[1]{>{\raggedright\arraybackslash}p{#1}}

\usepackage{enumitem}
\usepackage{float}
\nolinenumbers
\title{\nolinenumbers On Changes of Global Wet-bulb Temperature and Snowfall Regimes}
\authors{\nolinenumbers Sagar K. Tamang\correspondingauthor{Department of Civil, Environmental and Geo-Engineering, University of Minnesota-Twin Cities, Minneapolis, Minnesota}, Ardeshir M. Ebtehaj}
\affiliation{Saint Anthony Falls Laboratory and Department of Civil, Environmental and Geo-Engineering, University of Minnesota-Twin Cities, Minneapolis, Minnesota}
\extraauthor{\nolinenumbers Andreas F. Prein, Andrew J. Heymsfield}
\extraaffil{\nolinenumbers National Center for Atmospheric Research, Boulder, Colorado}
\email{\nolinenumbers taman011@umn.edu}

\abstract{\nolinenumbers To properly interpret the observed shrinkage of the Earth's cryosphere it is important to understand global changes of snowfall dominant regimes. To document these changes, three different reanalysis products of wet-bulb temperature together with observationally-based data sets are processed from 1979 to 2017. It is found that over the Northern Hemisphere (NH), the annual mean wet-bulb temperature has increased at a rate of 0.34\,$^\circ$C~per decade (pd) over land and 0.35\,\cpd over ocean, resulting in a reduction of the annual mean potential areas of snowfall dominant regimes by 0.52/0.34 million \kmpdend~over land/ocean. However, the changes in the Southern Hemisphere (SH) are less conclusive and more uncertain. Among the K$\ddot{\textrm{o}}$ppen-Geiger climate classes, the highest warming trend is observed over the NH polar climate regimes. Over studied mountain regions, the Alps are warming at a faster rate compared to the Rockies, Andes and High Mountain Asia. Due to such warming, potential snowfall areas over the Alps is reducing at 3.64\% pd followed by Rockies at 2.81 and HMA at 1.85\% pd. On average, these mountain ranges have lost 0.02 million~\kmpd of potential snowfall areas. The NH potential snowfall areas is retracting towards the North pole over the Central Asia and Europe at a rate of 0.45 and 0.7 degree pd. Furthermore, terrestrial regions over the NH including the Great Plains in the United States, Canadian provinces around the Hudson Bay, Central Siberian and Tibetan Plateaus, are losing as much as 4\% of the solid proportion of the annual precipitation amount pd.}  

\begin{document}
 \maketitle

\section{Introduction}
\label{S:1}

\par Snow and its meltwater play a crucial role in the global water and energy cycle. Snowpack stores freshwater in winter and releases it during the summer when it is needed the most \citep{Viviroli_2007,Wan2014ChangeChina}. In a warmer world, less proportion of winter precipitation falls as snow and winter snowpack melts earlier in spring, causing water shortages in summer \citep{Barnett_2007}. Climate projections indicate that a population of almost 2 billion people could be exposed to a high risk of decreased snow water supply in the next century \citep{Mankin_2015}. There are regional studies indicating declines of important snowpack reservoirs around the globe. Ground observations in the western United States show increased freezing elevations \citep{Ashfaq2013}, declined snow water equivalent \citep{Mote2005}, and earlier snowmelt runoff \citep{Rauscher2008}. \citet{marty2017recent} argued that the shrinkage of Alpine glaciers is due to a shorter duration of snow cover. A multi-decadal record of snow-cover satellite data \citep{Hall2002}, also indicates significant global shrinkage of snow-cover areal extent \citep{Brown2011}. 

Snowfall accumulation controls the mass balance of snowpack and glaciers. Regional studies report that the snowfall is decreasing over important mountainous regions of the world, including the Himalayas \citep{gusain2014winter,mir2015decline}, the Tibetan Plateau \citep{Wang2016Decrease2013}, Italian Alps \citep{valt2013climate}, the Pacific Coast Ranges \citep{Feng2007ChangesStates,Howat2005}, and the Tien-Shan mountains \citep{guo2015variation}. Despite significant progress in the understanding of regional changes in snowfall patterns, there is still a large gap in our understanding of the global changes in snowfall space-time distribution. Satellite precipitation estimates are promising to help close this knowledge gap in cold climate regions, which typically have sparse ground observations. However, unlike rainfall \citep{Bolvin2009,Huffman2009,Behrangi2016}, a sufficiently long and reliable record of satellite snowfall is still lacking.  The launch of the Global Precipitation Measurement (GPM, 2014-present) satellite \citep{Hou2014, Skofronick-Jackson_2017} is going to close this gap but its record length is too short to study long-term trends. Here, we explore whether and how existing long-term satellite precipitation and reanalysis data can be used to understand global snowfall change? 

Multi-sensor precipitation products such as the Tropical Rainfall Measurement Mission Multi-Satellite Precipitation Analysis \citep{huffman2013trmm} and Pentad Global Precipitation Climatology Project (GPCP) \citep{xie2003gpcp} provide records of cumulative precipitation obtained from a combination of rain gauges and retrievals from a series of spaceborne sensors. However, these products currently do not have any specific information on precipitation phase. The phase of precipitation can be inferred from near surface air temperature \citep{Kienzle2008ASnow,Dai2008TemperatureOcean} or the wet-bulb temperature \citep{zhong2018discriminating,Ding2014,Sims2015}. Numerous studies have demonstrated that wet-bulb temperature can capture the precipitation phase change with less uncertainty than the air temperature \citep{Ding2014} since it also accounts for the effects of the air moisture content. In particular, \citet{Sims2015} studied the uncertainty range defined as the difference between 10 and 90 percentiles of the snowfall conditional probability using 9700 stations over land and oceans from 1950 to 2007. This study found that uncertainties in characterizing precipitation phase changes are significantly reduced when wet-bulb temperature is used instead of air temperature. Their results show that precipitation is in solid form with more than 50\% probability when the near surface wet-bulb temperature is below \ang{1}\,C over land and \ang{1.1}\,C over oceans.

The goal of this paper is to understand and quantify the global changes in snowfall dominant regimes over different climate regimes and important mountainous regions of the world. In particular, the paper focuses on the K$\ddot{\textrm o}$ppen-Geiger arid-cold, cold and polar climate classes as well as four important mountain regions of the world including the Rockies, Alps, High Mountain Asia (HMA) and Andes. For our analysis, we use observational based precipitation estimates and 2\,m wet-bulb temperature from three third-generation reanalyses products. To quantify the global snowfall changes, the annual and seasonal changes in wet-bulb temperature, potential snowfall dominant areas, position of the snowfall transition latitudes as well as annual changes in Snowfall to Precipitation Ratio (SPR) were quantified and validated with the ground-based gauge observations by the National Climatic Data Center (NCDC).

\par The paper is organized as follows: Section 2 discusses the data sets and pre-processing tasks. The key measure of changes and used statistical approaches for trend identification are described in Section 3. Section 4 demonstrates and interprets the results while Section 5 concludes and discusses the findings and implications. Computation of the wet-bulb temperature using the approach by \citet{stull2016practical} and statistical measures used in the validation part are explained in the Appendix.

\section{Data and Preprocessing}
\subsection{Wet-bulb temperature}

The European Centre for Medium-Range Weather Forecast (ECMWF) interm reanalysis (ERA-Interim) \citep{Dee2011} uses four-dimensional variational (4D-Var) data assimilation of a vast amount of in situ and remote sensing observations. The data are provided at a spatial resolution of $\ang{0.125}$ every 6 hours from 1979 to present. There is research suggesting that the ERA-Interim performs well in the simulation of the surface air temperature \citep{mooney2011comparison,wang2012evaluation}, soil moisture \citep{peng2015evaluation} and surface wind velocity \citep{largeron2015can} compared to other reanalysis products.

JRA-55 is developed by Japanese Meteorological Agency \citep{KOBAYASHI2015TheCharacteristics} and also uses a 4D-Var data assimilation system. The data is made available at a spatial resolution of \ang{0.5625} every 6 hours from  1958 to present. Studies have shown that JRA-55 performs well for simulation of tropical cyclones \citep{murakami2014tropical} and rainfall diurnal cycle \citep{chen2014evaluation}.

NCEP-DOE R-2 \citep{Kanamitsu2002NCEP-DOER-2} uses a 3D-Var data assimilation system and provides data at a spatial resolution of \ang{2.5} every 6 hours from 1979 to present.  Studies suggest that NCEP-DOE R-2 performs well in capturing intense rainfall variability \citep{tesfaye2017evaluation}. 

\subsection{Precipitation}

The Pentad GPCP Version 2.2 provides multi-sensor estimates of 5-day surface global precipitation at a spatial resolution of \ang{2.5} from 1979 to 2016 \citep{xie2003gpcp,xie2011global}. This product is created by merging the Pentad Climate Prediction Center (CPC) Merged Analysis of Precipitation (CMAP) \citep{xie1997global} and the GPCP monthly multi-sensor precipitation product \citep{Adler2003The1979Present}. The Pentad CMAP dataset optimally combines gauge precipitation data from more than 6000 Global Telecommunication System stations together with precipitation estimates from Infra-red sensor on board the Geostationary Operational Environmental Satellite (GOES), Microwave Sounding Unit on Television Infrared Observation Satellite (TIROS), Special Sensor Microwave Imager (SSM/I) on board the Defense Meteorological Satellite Program (DMSP) satellites and Advanced Very High Resolution Radiometer (AVHRR) on board the National Oceanic and Atmospheric Administration (NOAA) operational sun-synchronous polar-orbiting satellites. A precipitation observation inter-comparison study over Europe has shown that GPCP has a tendency to overestimate precipitation amounts in flat regions and underestimate amounts in mountainous areas \citep{prein2017impacts}. Generally, there are large uncertainties in global precipitation observations, particularly in regions with low station density and snow dominated environments \citep{sun2018review}.

\begin{figure}[H]
\centering\includegraphics[width=1\textwidth]{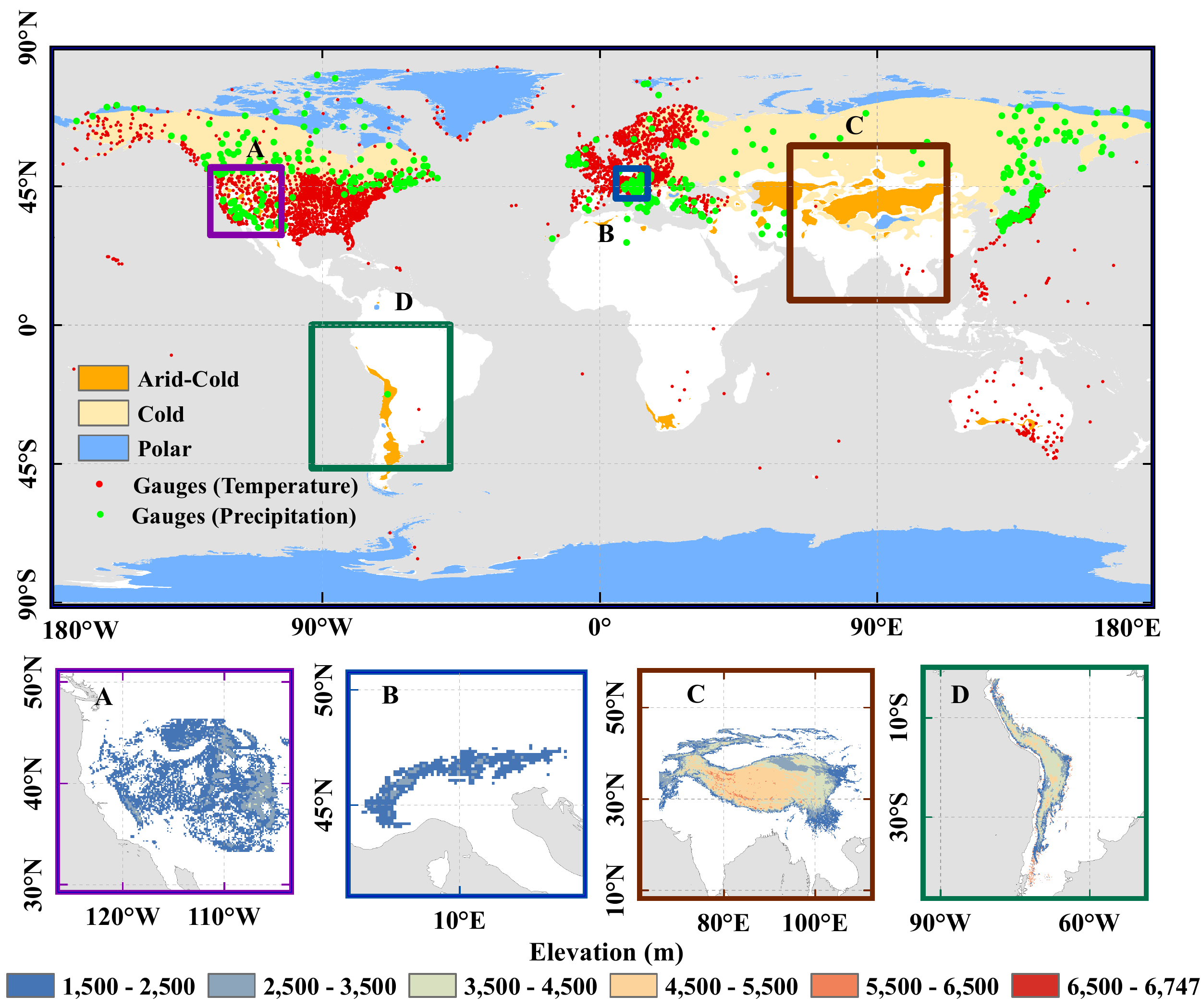}
 \caption{Location of the NCDC gauge stations (3579 for wet-bulb temperature and 427 for precipitation) used in the study for validation of the reanalysis based wet-bulb temperature and SPR from 2011 to 2015 (top row). The map shows three K$\ddot{\textrm{o}}$ppen-Geiger climate classes:  Arid-Cold (orange), Cold (yellow) and Polar (blue) \citep{peel2007updated}. Also shown are the (A) Rockies (B) Alps (C) HMA and (D) Andes mountains \citep{blyth2002mountain}. Elevation maps for the mountain regions are obtained from NOAA's digital elevation model at 1~km resolution (bottom row).}
\label{Fig.1}
\end{figure}

\subsection{Gauge Data}
Error covariance for the reanlaysis wet-bulb temperatures are calculated against gauge observations. Furthermore, the SPR obtained from the Pentad GPCP precipitation is validated against the ground-based precipitation gauge data. For this purpose, we used gauge data from the Global Surface Summary of the Day \citep[GSOD,][]{lott1998global,smith2011integrated}, which is provided by the NCDC from 2011 to 2015. For computation of the error covariance, 3579 NCDC stations with $\sim$4.1 million station days (data from one station in one day) of wet-bulb temperature were used (Fig.~\ref{Fig.1}), while only 859 station years could be utilized for annual validation of the SPR. The reason is that all stations with more than 10 days per year of missing information on precipitation amount and phase were not utilized.

\section{Methodology}

\subsection {Reanalysis Ensemble Mean}
\label{section:3.6}
Multiple reanalysis data can be considered as an ensemble realization of the underlying variable of interest and can be integrated for reducing the uncertainty of inference \citep{hagedorn2005rationale,solman2016climate}. In this study, we use the maximum likelihood (ML) estimator of the reanalysis products. To that end, we assume that the 2~m wet-bulb temperature $T_{wb}^{t,m}$ by reanalysis product $m$ at time $t$ is related to the ground truth wet-bulb temperature $T_{wb}^{t}$ as follows:

\begin{equation}
    T_{wb}^{t,m} = T_{wb}^{t} + \epsilon_m, \quad \quad \quad \epsilon_m \sim \mathcal{N}(0,\sigma_m^2)
\end{equation}

\noindent where, the reanalysis error $\epsilon_m$ are independent zero-mean normally distributed random variables with variance $\sigma_m^2$. Therefore, the likelihood function can be obtained as $\mathcal{L}(T^{t,m}_{wb}|T^t_{wb})\propto \Pi_{m=1}^3\exp{\left[-(T^{t,m}_{wb}-T^t_{wb})^2/2\,\sigma_m^2\right]}$ for which the ML is the estimate of the reanalysis wet-bulb temperature $\hat{T}_{wb}^t$ is:

\begin{equation}
    \hat{T}_{wb}^t=\argmin_{T_{wb}^t} \: \sum_{m=1}^{3} \frac{(T_{wb}^{t,m}-T_{wb}^t)^2}{\sigma_m^2},
    \label{eqn:optim}
\end{equation}

\noindent which leads to 
\begin{equation}
    \hat{T}_{wb}^t = \sum_{m=1}^{3} w_m\,T_{wb}^{t,m},
\end{equation}
\noindent where, $w_m=\sigma_m^{-2}\,\left(\Sigma_m\sigma_m^{-2}\right)^{-1}$ and variance of the ML estimator is given by $(\sum_m \sigma_m^{-2})^{-1}$. Because of the Gaussian assumption, the ML estimate is equivalent to an ensemble mean that is weighted based on the variance of the uncertainty of each product. Note that in the above formalism, we assumed that the reanalysis data are unbiased, which is a reasonable assumption as will be shown later.

\subsection{Potential Snowfall Area}

A reanalysis field of wet-bulb temperature at pixel-level $(i,\,j)$ is denoted by $T_{wb}(i,\,j)$. Here, the potential snowfall dominant regimes are defined as the interior area of the set $\mathcal{S}$ of all global pixels over which $T_{wb}(i,\,j)\leq T_{wb}^s$, where $T_{wb}^s$ denotes the temperature threshold below which the probability of snowfall is above 50\%.
In this study, we set $T_{wb}^s=1^{\circ}$C and $1.1^{\circ}$C over land and oceans respectively \citep{Sims2015}, to define the boundaries $\partial \mathcal{S}$ of the set. Throughout, we use daily wet-bulb temperature data and their ensemble mean, to infer seasonal and annual changes of potential snowfall areas and other measure of changes that will be explained in the subsequent sections. 

\subsection{Snowfall to Rainfall Transition Latitudes}

The analysis of the potential snowfall areas provide a bulk quantitative indication on how snowfall dominant regimes are shrinking or expanding over time; however, another key question is: where, and to what extent, the snowfall dominant regimes are changing into rainfall dominant regimes? The location and movement of the boundary of potential snowfall areas capture the regions that are experiencing the most significant interannual variability of snowpack water storage and related hydrologic response. As explained before, this freezing boundary $\partial \mathcal{S}$, can be defined as the contour of $T_{wb}^s$ that separates potential snowfall dominant regions from the rainfall dominant regimes. However, because of the isolated temperature islands, quantifying the rate of change is not straightforward. To overcome this challenge, the annual zonal mean transition latitudes are calculated over longitudinal slices of 15$^{\circ}$ that enclose sector areas equal to the annual mean of potential snowfall area within that particular slice.

\subsection{Snowfall to Precipitation Ratio}

To obtain a more realistic measure of snowfall changes, we investigate the snowfall to precipitation ratio (SPR) by combining the ensemble mean reanalysis wet-bulb temperature data with the Pentad GPCP precipitation. The ensemble mean of reanalysis wet-bulb temperatures is computed daily at the spatial resolution of ERA-Interim \ang{0.125}; however, cumulative Pentad GPCP precipitation is available every 5 days at a spatial resolution of \ang{2.5}. Therefore, the Pentad GPCP data is first mapped onto the spatial resolution of ensemble mean through nearest neighbor interpolation to avoid any loss or addition of spurious information. Then, the relative frequency of snowfall occurrence for each pixel is defined as the ratio of the number of days when the daily wet-bulb temperature is below the snowfall threshold to the 5-days temporal resolution of the Pentad GPCP data. The computed relative frequency is then multiplied with the precipitation amount to obtain an estimate of the snowfall amount. Annual SPR is finally obtained by dividing cumulative snowfall amount by the annual cumulative precipitation amount as follows:
 
 \begin{equation}
     \text{SPR}(i,j) = \frac{\sum_{t=1}^{n_p} f_s^{t}(i,j)\,P(i,j)}{P(i,j)},
 \end{equation}
where the relative frequency of snowfall occurrence is $f_s^{t}(i,j)=\frac{1}{T}\sum_{t=1}^{T} \mathbf{1}_{\mathcal{S}}\left[T_{wb}^t(i,\,j)\right]$, $P(i,j)$ represents the pixel-level cumulative precipitation during the Pentad GPCP temporal resolution and $n_p$ denotes the number of 5 days precipitation data per year.

\subsection{Trend Analysis}

We use the non-parametric Theil-Sen method \citep{Theil1950AI.,Sen1968EstimatesTau} for computing the magnitude of linear trends. The Theil-Sen method computes the trend by taking the median of the slopes of all possible lines that are fitted to pairs of sample points. This method does not require any parametric assumption about the probability distribution of the samples and exhibits higher degree of accuracy than the ordinary least square (OLS), in the presence of heteroscedasticity \citep{wilcox2010fundamentals}. Additionally, since this approach relies on the median of the slopes, the estimated trends are more robust to observational outliers than the OLS, which approximates the mean value of the trends \citep{matouvsek1998efficient,wilcox2010fundamentals}. 

\sloppy For brevity, here we explain the method for the hemispherical wet-bulb temperatures only. To that end, let us assume that $\mathbf{T}_{wb}=\left\{ \overline{T}_{wb}^{1},\,\overline{T}_{wb}^{2},\ldots,\,\overline{T}_{wb}^{n}\right\}$ represents annual time series of hemispheric mean wet-bulb temperatures for year \textbf{\emph{y}}$=\{y^1,\,y^2,\ldots,\,y^n\}$. The Theil-Sen estimate of the linear slope ($\beta$) is defined as follows:
%============================Equation 1
\begin{equation}
\beta = \text{median}\bigg( \frac{\overline{T}_{wb}^j-\overline{T}_{wb}^i}{y^j-y^i}\bigg ), \;\quad i= 1, 2,\ldots, n-1, \;\quad j= 2, 3,\ldots, n, \;\quad j>i.
\label{eqn:TS}
\end{equation}
%======================================
Numerous tests have been examined to quantify the statistical significance of the  Theil-Sen estimator such as the parametric {\it t}-test \citep{student1908probable} and non-parametric Mann-Kendall (MK) test \citep{Mann1945NonparametricTrend,kendall1948rank}. Here we adopt the bootstrap MK (BS-MK, \citep{douglas2000trends}) test as \citet{yue2004comparison} showed that this method has higher probability of correct rejection of the null hypothesis for linear trend detection of non-Gaussian data structure, among other commonly used tests. In summary, the MK test computes the following test statistic:
%============================Equation 2
\begin{equation}
 S= \sum_{k=1}^{n-1} \sum_{j=k+1}^{n} \text{sgn}(\overline{T}_{wb}^j-\overline{T}_{wb}^k), 
\label{eqn:S Statistic}
\end{equation}
%======================================
where $j>k$ and ${\rm sgn}(\cdot)$ refers to the signum function. Positive (negative) values of $S$ imply a positive (negative) trend in the the time series. Under the null hypothesis, \citet{kendall1948rank} showed that the test statistic $S$ is asymptotically a zero-mean normally distributed random variable with variance  $\mathbb{V}ar(S)=[n(n-1)(2n+5)-\sum_{i=1}^{q}t_i(t_i-1)(2t_i+5)]/18$,
where, $q$ is the total number of groups of same observations or ties and $t_i$ is the number of observation in the $i^{th}$ tied group.

Thus the standard test statistic ($Z$) is defined as follows:
\begin{equation}
Z=
\begin{cases}
	\displaystyle \frac{S-1}{\sqrt[]{\mathbb{V}ar(S)}} \quad \textnormal{if} \quad S>0 \\
	\displaystyle \quad \;0 \quad \; \quad \textnormal{if} \quad  S=0\\
    \displaystyle \frac{S+1}{\sqrt[]{\mathbb{V}ar(S)}} \quad \textnormal{if} \quad S<0 \\
\end{cases},
\label{Eqn:Z}
\end{equation} 
The null hypothesis can be rejected if $|Z|\ge Z_{1-\alpha/2}$, at $\alpha$ level of significance.

\par The asymptotic null distribution of the MK test statistic $S \stackrel{a}{\sim} \mathcal{N}(0,\mathbb{V}ar(S))$ is valid under the assumption of serial independence \citep{vonStorch1995MisusesClimate}.
To formally account for the effects of serial dependence the Pre-Whitening \citep{vonStorch1995MisusesClimate} and Trend Free Pre-Whitening \citep{yue2002influence} approaches have been proposed. \citet{yue2004mann} showed that the presence of positive (negative) serial correlation in the data inflates (deflates) the variance of MK test statistic and thus proposed a variance-correction method. Additionally, block bootstrap approaches \citep{kundzewicz2000detecting,Onoz2012BlockData} have been suggested to approximate directly the null distribution of the MK test statistic through resampling, without removing the serial dependence of the data. \citet{Khaliq2009IdentificationRivers} compared the performance of the explained methods and found that the pre-whitening methods are conservative in identifying significant trends while both variance correction and block bootstrap methods perform well for dependent time series.

The variants of the block bootstrap method \citep{kunsch1989jackknife,carlstein1986use,kunsch1989jackknife,liu1992moving,politis1994stationary} are an extension to the original bootstrap inference approach \citep{Efron1979BootstrapJackknife} for approximating the sample distribution of a statistic in serially dependent data sets. This method reconstructs the bootstrap samples through resampling of data blocks beyond which, the dependent structure of the data becomes negligible. Here, we confine our consideration to the classic moving block bootstrap \cite[MBB,][]{liu1992moving}, which has been applied and tested successfully for significance analysis of the MK estimates of linear trends \citep{Khaliq2009IdentificationRivers,Onoz2012BlockData}. 

\par Specifically, let us assume that $S$ is the the MK test statistic of the original annual time series $\mathbf{T}_{wb}=\left\{ \overline{T}_{wb}^{\,1},\,\overline{T}_{wb}^{\,2},\ldots,\,\overline{T}_{wb}^{\,n}\right\}$. Given the time series, a serial correlation length $r$ at $\alpha$ significance level is computed and the length of the block is set to $l=r+1$. The time series $\mathbf{T}_{wb}$ is then divided into $N=n-l+1$ overlapping blocks $\xi_i:=\left\{\overline{T}_{wb}^{\,i},\ldots,\overline{T}_{wb}^{\,i+l-1}\right\}$, with probability of occurrence equal to $1/N$, where $i=1,\ldots,N$ and $l/n\rightarrow 0 $ as $n,l\rightarrow \infty$. A number of $k=\lceil n/l \rceil$ blocks  $\left\{\xi^{*b}_1,\xi_2^{*b},\ldots,\xi_k^{*b}\right\}$ are sampled with replacement from $\xi_1,\xi_2,...,\xi_N$ and concatenated to reconstruct $B$ bootstrap pseudo time series as $\overline{T}_{wb}^{*b}=\{\xi^{*b}_1,\xi_2^{*b},\ldots,\xi_k^{*b}\}$, where $b=1,2,\ldots,B$. Bootstrap empirical distribution of MK test statistic ($S^*$) is obtained from the pseudo bootstrap time series $\overline{ T}_{wb}^{*b}$ and the significance of trend is finally computed by applying a two-tailed hypothesis test. If the MK test statistic $S$ of the original time series is higher than the 97.5 percentile or lower than 2.5 percentile of the empirical distribution of test statistic $S^*$, the hypothesis, that there is no trend in the data, is rejected. Throughout, for computation of the empirical distribution of MK test statistic, we set $B=3000$.

Throughout, the trends at significance level $\alpha$ are reported as $\beta_{\alpha}\:(\beta_{min}-\beta_{max})$, where $\beta_{\alpha}$ denotes the trend of ensemble mean with an appropriate unit and values in parentheses denote the minimum and maximum value of detected trends by the three reanalysis products. The reported changes without a subscript are insignificant at $\alpha=0.05$.

\section{Results}

\subsection{Changes in Wet-bulb Temperature}
\label{section:4.1}
In this subsection the spatial variability and zonal mean values of changes in the global wet-bulb temperature are characterized and discussed for the reanalysis products. Then, characterizing the error variance of the three reanalysis products, the space-time variability of the ensemble mean is analyzed and discussed over the aforementioned climate regimes and mountainous regions of the world. Finally, the temporal changes of the annual and seasonal mean wet-bulb temperature are quantified on a hemispherical scale over land and oceans. 

\begin{figure}[H]
\centering\includegraphics[width=1\textwidth]{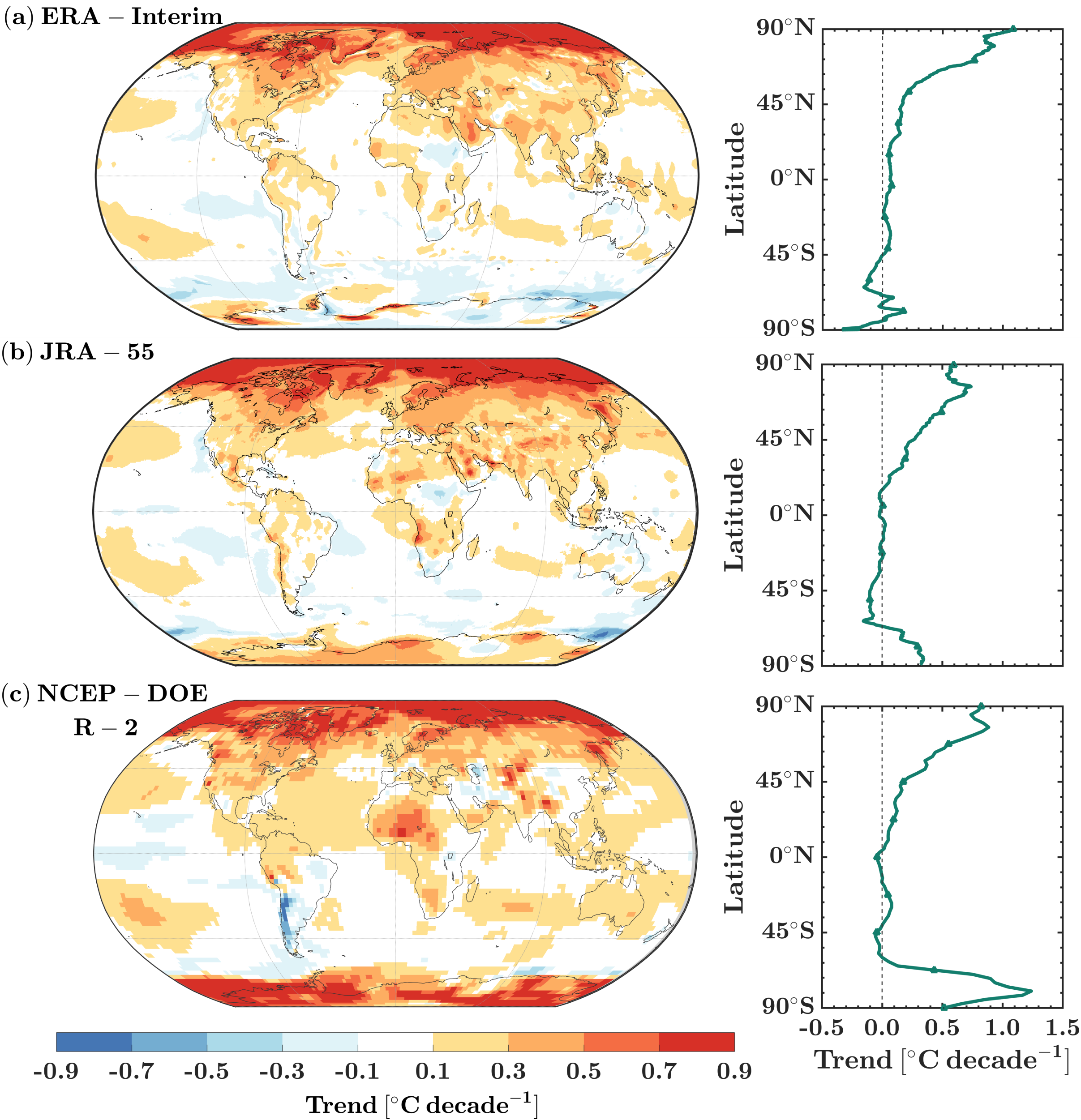}
\caption{Trend of annual mean wet-bulb temperature (left) and zonal trend of annual mean wet-bulb temperature (right) for the used reanalysis products from 1979 to 2017.}
\label{Fig.2}
\end{figure}

Fig.\ref{Fig.2} shows the spatial distribution of the trends in annual mean (left) and the trend of the zonal mean (right) wet-bulb temperature during the study period. Over the NH, there is a good agreement between the three data sets, especially with respect to the detected positive trends over cold and polar climate regimes above the arctic circle. In particular, large areas of the Canada's tundra and boreal forests, Midwestern United States, the Greenland ice sheet, Northern Europe, the Central Siberian Plateau, and the Scandinavian Peninsula have experienced a warming trend above 0.3~\cpd. Over higher latitudes (70--\ang{90}N), zonal mean trends are consistently positive and larger than 0.5~\cpd  in all three reanalysis products (Fig.~\ref{Fig.2}, right column). Some disagreements can be observed over lower latitudes, especially over the Iranian Plateau, Indian Peninsula and east Asia, where only ERA-interim and JRA-55 indicate a coherent warming trend. The disagreement is more apparent over the western Africa and across the north Atlantic Ocean, where only the NCEP-DOE shows a coherent warming trend. Over the SH, there is large disagreement between the reanalysis data sets. The NCEP-DOE indicates that the annual mean wet-bulb temperature is decreasing coherently over the Andes and increasing over Australia, whereas these trends are not apparent in other reanalysis products. The disagreement is most pronounced over Antarctica where ERA-Interim does not show any spatially organized trend while a positive trend is seen in the other data sets, especially in the NCEP-DOE. This warming trend is more significant and reaches almost 0.6~\cpd near the Antarctica with a peak around \ang{80}S.

\begin{figure}[h]
\centering\includegraphics[width=1\textwidth]{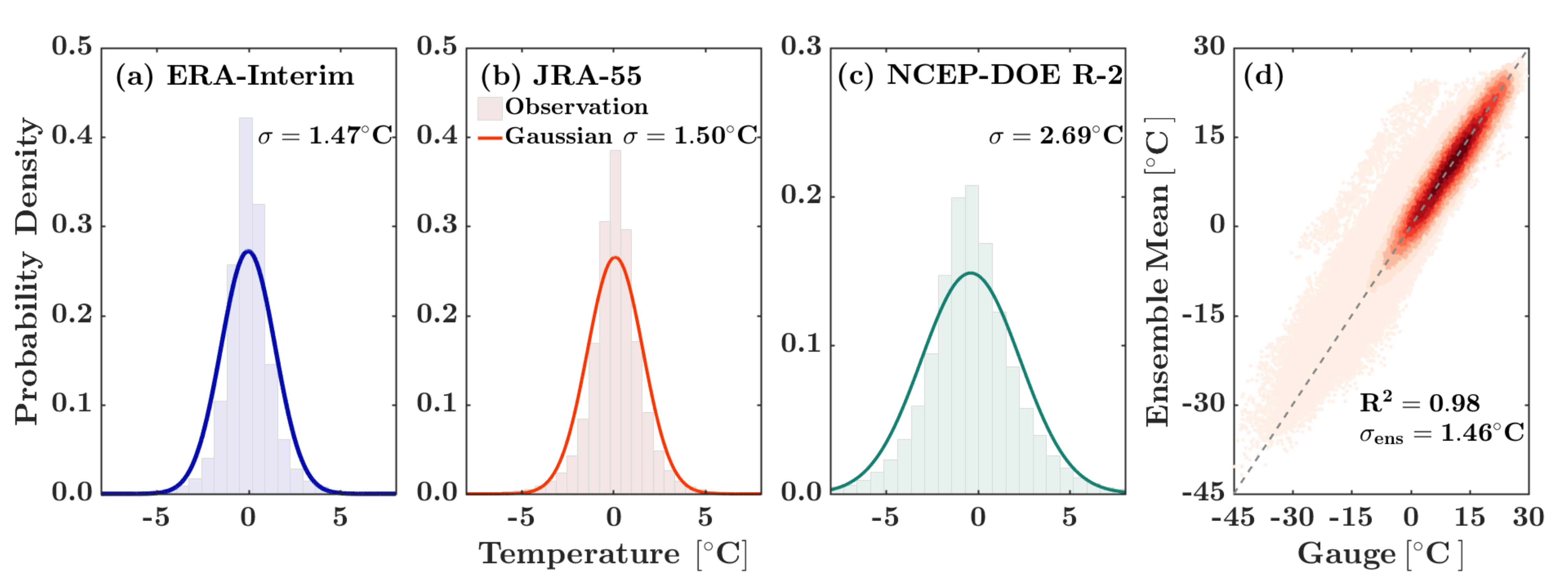}

\caption{Probability histogram of error in the wet-bulb temperature from the (a) ERA-Interim, (b) JRA-55, (c) NCEP-DOE R-2 within the validation period (2011 to 2015) (d) scatter density plot of the ensemble mean versus the NCDC GSOD gauge data.}
\label{Fig.3}
\end{figure}

\subsection{Ensemble Mean Wet-bulb Temperature}
As previously explained, we use ground-based gauge stations from 2011 to 2015 to understand the reanalysis error and validate some of the results. The error distribution of the reanalysis wet-bulb temperatures is shown in Fig.~\ref{Fig.3} (a--c). The errors, for each reanalysis product, were computed using more than 4.1 million data points from 3579 NCDC gauges stations (Fig.~\ref{Fig.1}) on a daily basis. The reanalysis data are almost unbiased with respect to the areas that are densely populated by the gauges. Fitted Gaussian distributions have standard deviation of \ang{1.47}, \ang{1.5} and \ang{2.69}~C for the ERA-Interim, JRA-55 and NCEP-DOE R-2, respectively. As shown in Fig.~\ref{Fig.3}~d, the ensemble mean wet-bulb temperature compares well with the gauge data as the the coefficient of determination ($R^2$) reaches to 0.98 and the ensemble error standard deviation is reduced to \ang{1.46}~C. It is worth nothing that, when the outliers (4.4\% of the data) are removed, using the Median Absolute Deviation (MAD) method, the ensemble error standard deviation reduces to \ang{1.01}~C, which is close to the theoretical standard deviation of \ang{0.96}~C by the ML estimator. 

\begin{figure}[H]
\centering\includegraphics[width=1\textwidth]{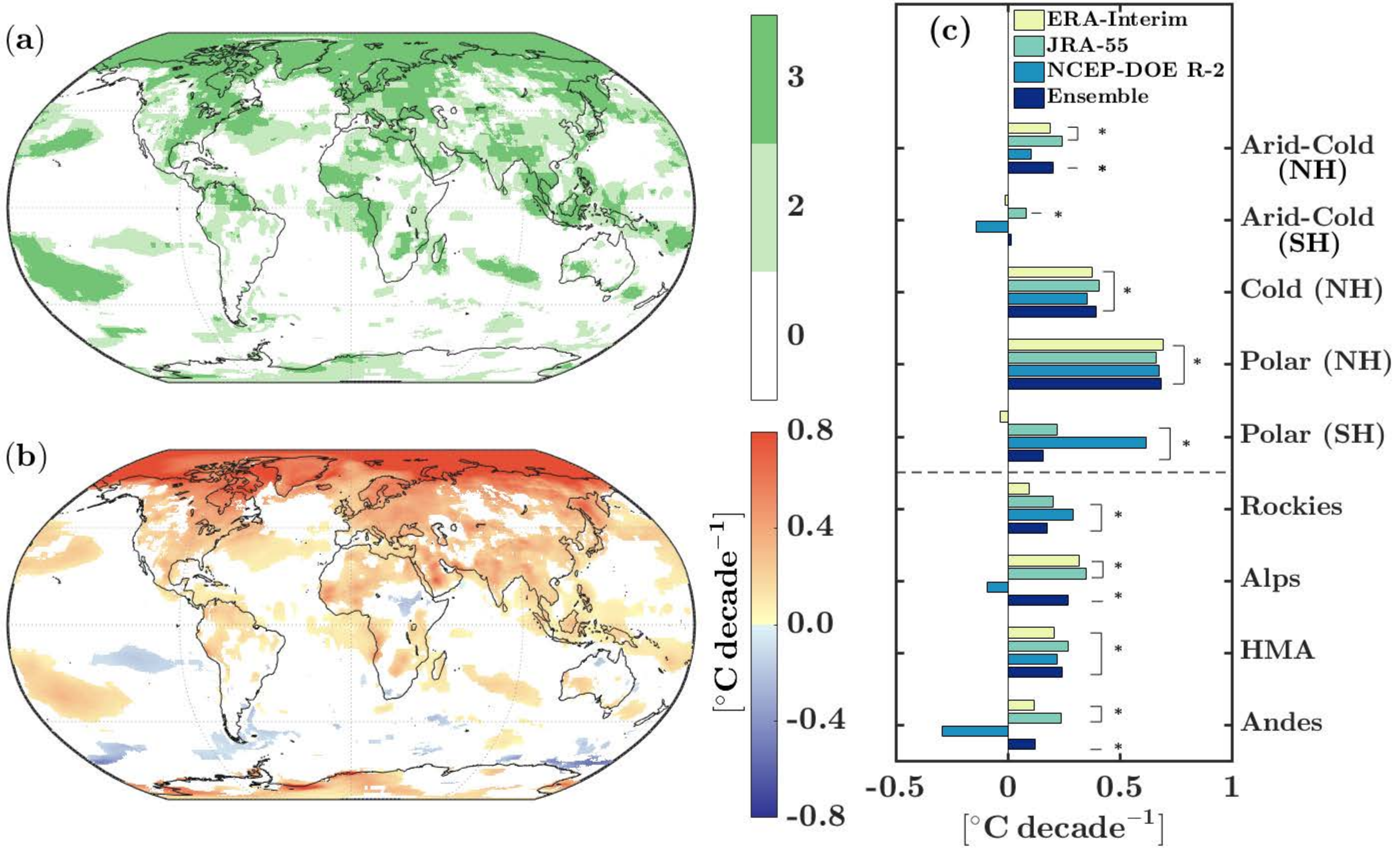}\\
\centering\includegraphics[width=0.75\textwidth]{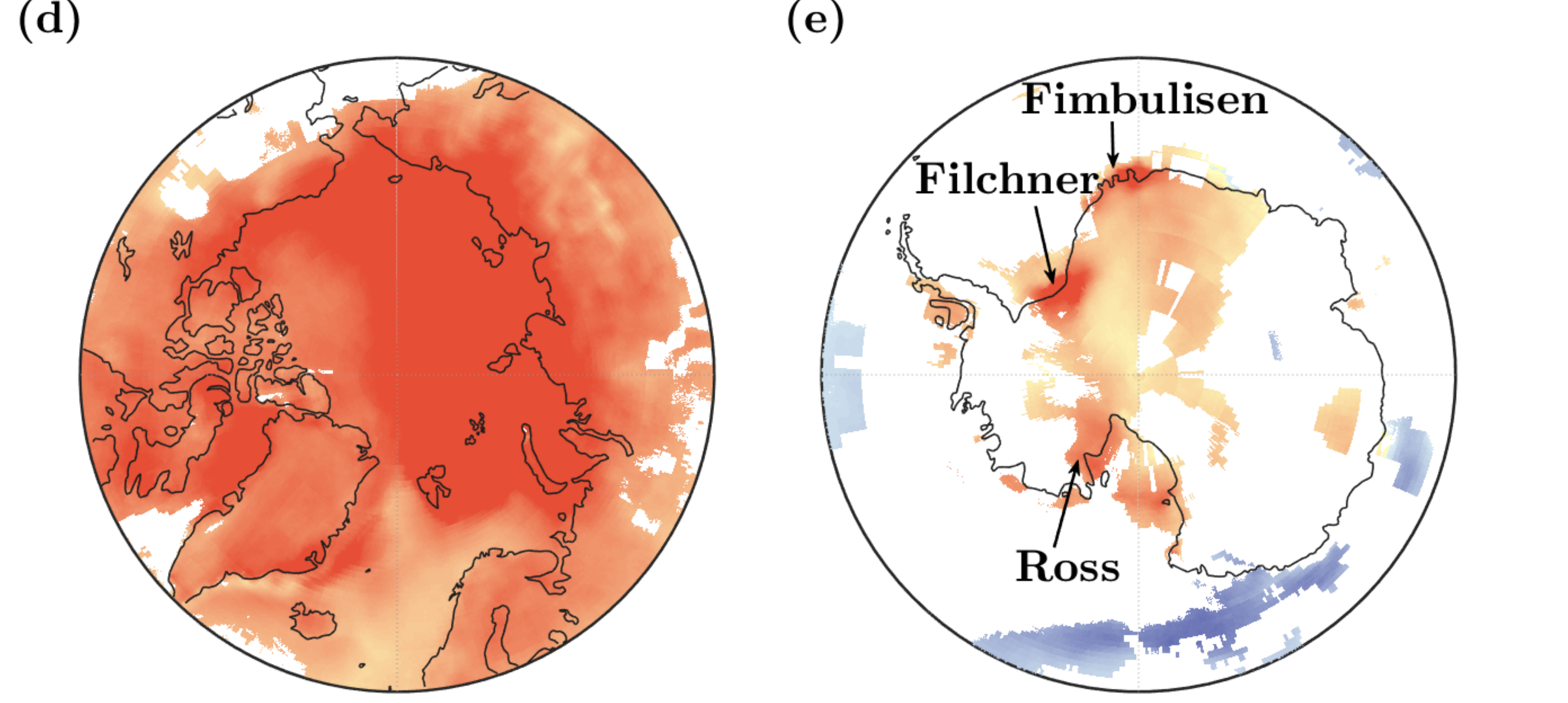}
\caption{Spatial distribution of the agreement in significant trends ($\alpha=0.05$) between the three reanalysis products (a), spatial distribution of the trends in the wet-bulb temperature (b) and its ensemble mean over the K$\ddot{\textrm{o}}$ppen-Geiger climate classes and four mountain regions (c). In the bar plots, significant trends are labeled by an asterisk. The trends are also shown in a polar projection system over the Arctic (d) and Antarctica (e).}
\label{Fig.4}
\end{figure}

In order to better understand the spatial pattern of the trends, binary masks of significant warming and cooling trends ($\alpha=0.05$) are created for each reanalysis product. The binary masks are then overlaid to identify the areas where the reanalysis data are in agreement for the computed trends (Fig.~\ref{Fig.4}\,a). The trend of the annual ensemble mean wet-bulb temperatures is computed and mapped onto the ERA-Interim 0.125$^\circ$ grid, using nearest neighbor interpolation, over the areas where there is a majority agreement (at least two reanalyses agree in trends) (Fig.~\ref{Fig.4}\,b). All products agree that there has been a significant trend of more than 0.4~\cpd over high latitudes above $60^\circ$\,N in the NH; however, only parts of Western Antarctica experiences robust positive trend.

\begin{figure}[H] 
\centering\includegraphics[width=1\textwidth]{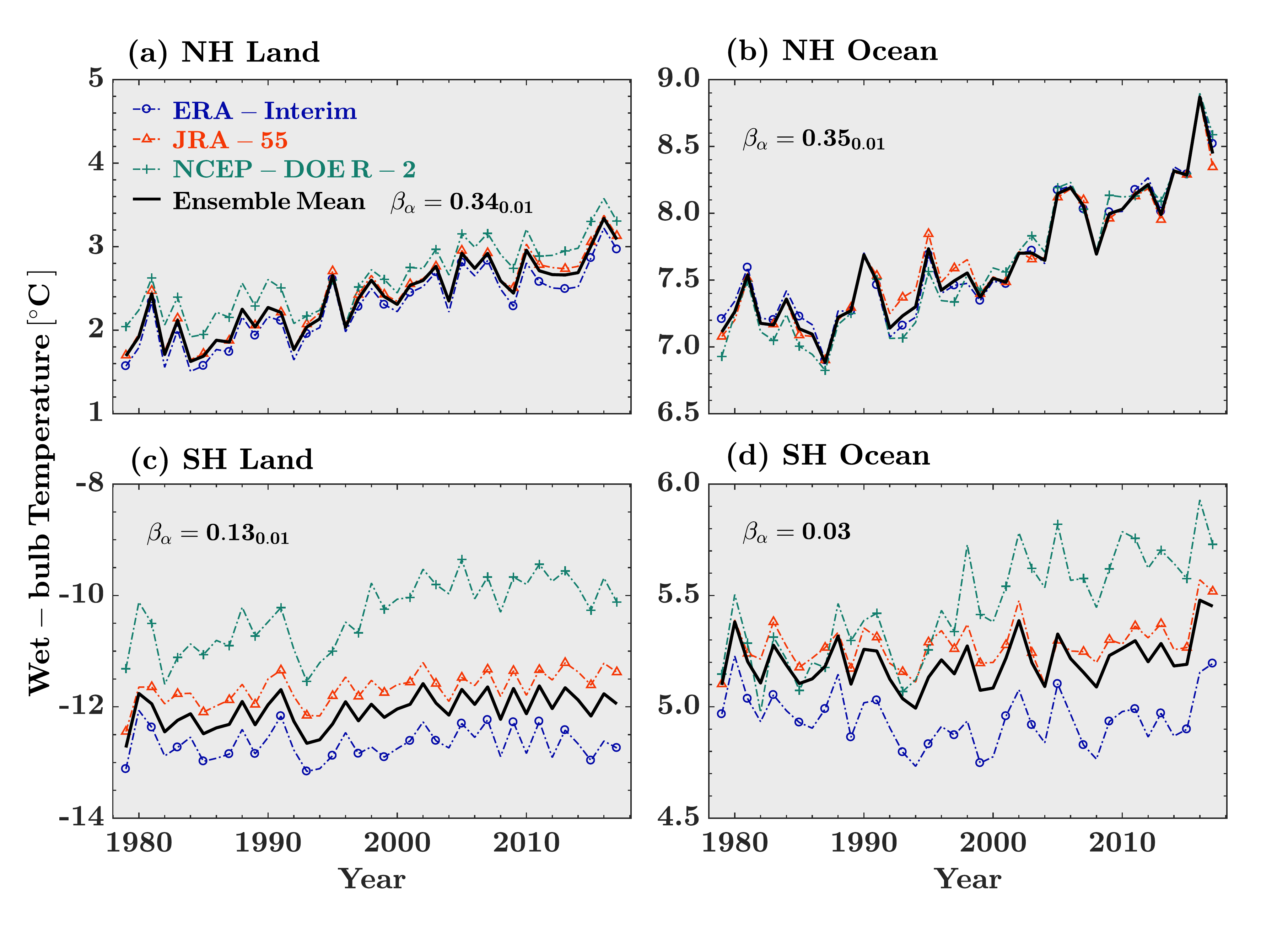}
\caption{Time series of annual mean wet-bulb temperature for the NH land (a) and oceans (b) and the SH land (c) and oceans (d) from 1979 to 2017 for the three reanalysis products and their ensemble mean.}
\label{Fig.5}
\end{figure}

Over the NH lands, the Midwest and Northeast regions of the U.S. are showing positive trends in all models. The positive trends extend over the European continent and Western Eurasia (except the Iberian Peninsula). Large parts of  Southeast Asia, the Arabian Peninsula, and West and North Africa are experiencing significant warming trend, which are largely below 0.4~\cpd. Over the SH lands, areas of positive trends are less coherent. However, Western Africa, parts of Amazon basin and South East Asia are experiencing a spatially coherent warming trend. The majority of positive trends agree only over parts of Antarctica. Significant trends are observed over the Queen Maud Land and coasts of the Ross Dependency. It is important to note that positive trends above 0.4~\cpd are concentrated over the Filchner, the Fimbulisen and the Ross Ice Shelves. We want to point out that the wet-bulb temperature is mostly below the snowfall threshold over Antarctica and thus this warming trend may not directly lead to reduction of snowfall accumulation processes but could accelerate the ablation processes. 

\begin{table}[t]
\centering
\caption{Annual and seasonal changes of the hemispheric mean wet-bulb temperature over the land and oceans. The values in bold fonts are significant at $\alpha$~= 0.05. Due to the absence of information on December of 1978, the data over the NH winter (SH summer) were not analyzed in that year.}
\begin{tabular}{R{1.7cm}L{1.1cm}L{1.1cm}L{1.1cm}L{1.1cm}L{1.1cm}L{1.1cm}L{1.1cm}L{1.1cm}}
\toprule
Seasons & \multicolumn{2}{c}{ERA-Interim} & \multicolumn{2}{c}{JRA-55} & \multicolumn{2}{c}{NCEP-DOE R-2}&
\multicolumn{2}{c}{Ensemble}\tabularnewline
 & Land & Ocean & Land & Ocean & Land & Ocean & Land & Ocean\\
\midrule
{\bf Northern} &&&\\
Annual & $\bf +0.33$ & $\bf+0.35$ & $\bf+0.35$& $\bf+0.34$ & $\bf+0.32$ & $\bf+0.41$& $\bf+0.34$& $\bf+0.35$
\\
  Winter & $\bf+0.23$ & $\bf+0.51$ & $\bf+0.26$& $\bf+0.44$ & $\bf+0.24$ & $\bf+0.49$&
  $\bf+0.25$&
  $\bf+0.47$
\\
Spring & $\bf+0.37$ & $\bf+0.37$ & $\bf+0.37$& $\bf+0.32$ & $\bf+0.36$ & $\bf+0.38$&
$\bf+0.37$&
$\bf+0.36$
\\
 Summer&  $\bf+0.25$ & $\bf+0.12$ & $\bf+0.28$& $\bf+0.14$ & $\bf+0.24$ & $\bf+0.16$&
 $\bf+0.26$&
 $\bf+0.14$
 \\
 Fall& $\bf+0.38$ & $\bf+0.44$ & $\bf+0.37$& $\bf+0.49$ & $\bf+0.35$ & $\bf+0.61$&
 $\bf+0.37$&
 $\bf+0.49$\\
 \midrule
{\bf Southern} \\
Annual & $+0.04$ & $-0.02$ & $\bf+0.15$& $+0.03$ & $\bf+0.41$ & $\bf+0.16$&$\bf+0.13$&$+0.03$
\\
  Winter & $+0.01$ & $+0.02$ & $+0.15$& $+0.03$ & $\bf+0.53$ & $\bf+0.31$&
  $+0.12$&
  $\bf+0.07$
\\
 Spring & $\bf+0.15$ & $+0.04$ & $\bf+0.29$& $\bf+0.06$ & $\bf+0.56$ & $\bf+0.21$&$\bf+0.26$&$\bf+0.07$
\\
  Summer & $-0.06$ & $-0.04$ & $+0.14$& $\bf+0.05$ & $+0.13$ &
  $-0.04$&
  $+0.06$&
  $+0.00$
 \\
  Fall & $+0.01$ & $-0.06$ &
  $+0.11$&
  $-0.02$ & $\bf+0.33$ & $\bf+0.17$&
  $+0.09$&$-0.01$
\\
\bottomrule
\end{tabular}
\label{Tab:1}
\end{table}

The results in Fig.~\ref{Fig.4}\,c indicate that the changes are significant for all climate regimes expect over the SH's Arid-Cold. Specifically, the ensemble mean shows the highest warming rate over the NH's polar climate regime at $0.69_{0.05}$ (0.66-0.69)~\cpdend, followed by NH's areas with a cold and arid-cold climates at $0.39_{0.01}$ (0.36--0.41) and $0.20_{0.01}$ (0.11--0.24)~\cpd, respectively. The majority of the reanalysis products indicate that there has been a warming trend over the mountainous regions studied, among which the Alps are experiencing the highest positive trend at $0.27_{0.01}$ (-0.09--0.35)~\cpd followed by the HMA at $0.24_{0.01}$ (0.21--0.27) and Rockies at $0.18_{0.05}$ (0.10--0.29)~\cpd. 

Fig.~\ref{Fig.5} shows the time series of the NH and SH annual mean wet-bulb temperature over land and oceans, while Table~\ref{Tab:1} reports the annual and seasonal changes. All reanalysis products agree over the NH both in terms of the trends and their annual mean values, whereas over the SH, the results show large uncertainties. The highest mean wet-bulb temperature is observed in 2016 for the NH, both over land and oceans among all reanalysis products, whereas there is not such an agreement over the SH. Trend analysis suggests that over the NH, the annual ensemble mean wet-bulb temperature is rising over land at $0.34_{0.01}$ (0.32--0.35)~\cpd and oceans at $0.35_{0.01}$ (0.34--0.41). Over the NH, among all seasons, warming is the highest during the fall both over land at $0.37_{0.05}$ (0.35--0.38)~\cpd and ocean at $0.49_{0.05}$ (0.44--0.61) and the lowest overland during the winter at $0.25_{0.01}$ (0.23--0.26) and over ocean in summer at $0.14_{0.01}$ (0.12--0.16)~\cpdend (see Table~\ref{Tab:1}). Similar to the NH, annual mean wet-bulb temperature over the SH is rising at rate of $0.13_{0.01}$ (0.04--0.41)~\cpd and 0.03 (-0.02--0.16) over land and oceans, respectively. The SH spring manifests the highest warming rates over both land and oceans with the rate of $0.26_{0.01}$ (0.15--0.56)~\cpd and $0.07_{0.01}$ (0.04--0.21), respectively; however, summer manifests the lowest warming rate over land at 0.06 (-0.06--0.14) and ocean at 0.00 (-0.04--0.05)~\cpdend. It is important to note that, over the SH oceans, an insignificant cooling trend is observed during the fall.

\begin{figure}[H]
\centering\includegraphics[width=1\textwidth]{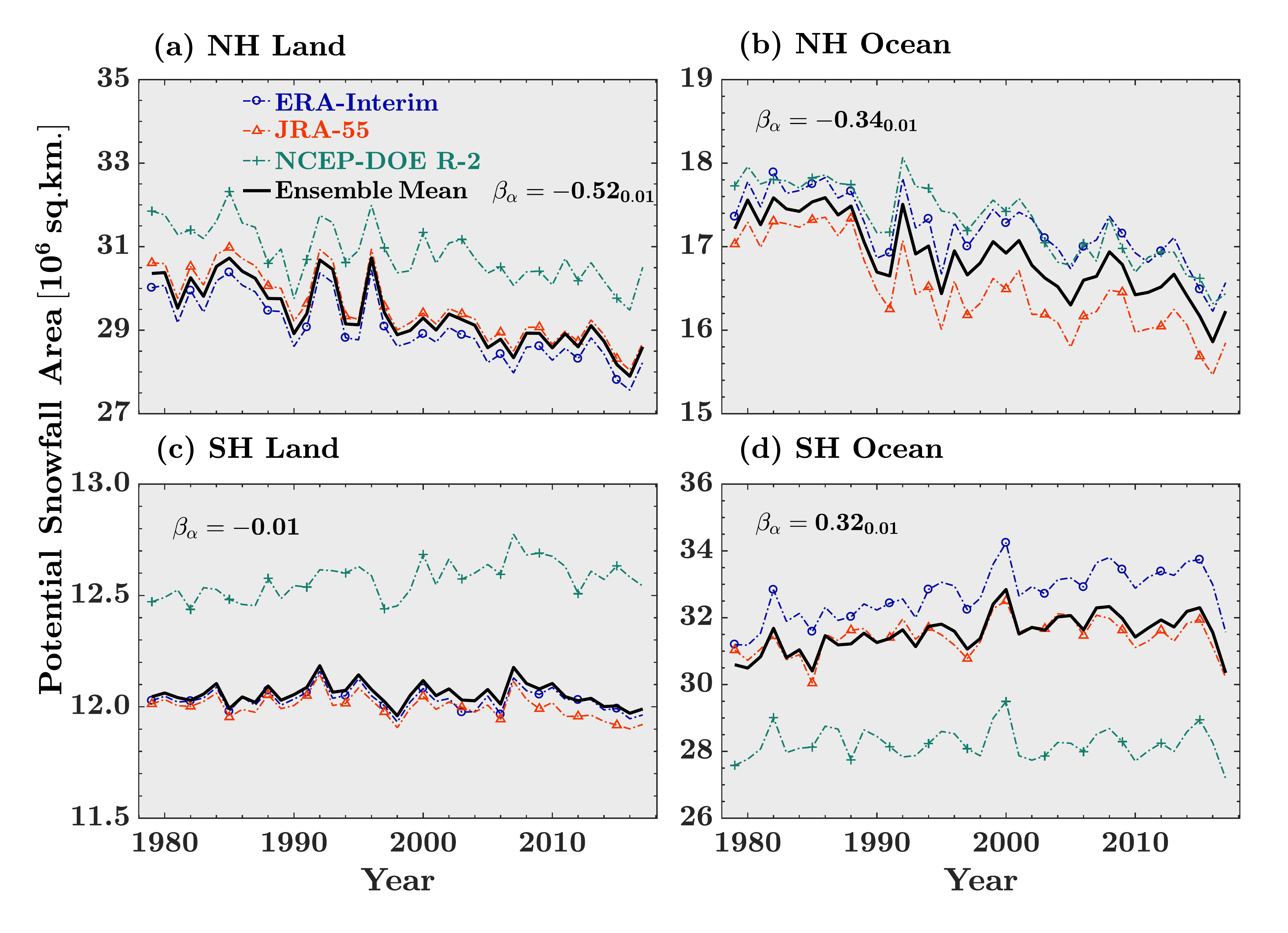}

\caption{Time series of the annual mean of the potential snowfall areas for the NH land (a) and oceans (b) and the SH land (c) and oceans (d) from 1979 to 2017 for the three reanalysis products and their ensemble mean.}

\label{Fig.6}
\end{figure}

\subsection{Changes in Potential Snowfall Areas}
\label{secion:4.2}

The annual time series of the hemispherical mean values of the potential snowfall areas in million~\kmpd are shown in Fig.~\ref{Fig.6} and the seasonal values are also reported in Table~\ref{Tab:2}. Over the NH land, potential snowfall areas are decreasing annually at a rate of $0.52_{0.01}$ (0.44-0.56) million~\kmpdend, which is more than the NH oceans $0.34_{0.01}$ (0.28--0.39). Seasonal analysis suggests that the shrinkage rate is largest ($0.81_{0.01}$ (0.58--0.86) million~\kmpd) during spring over the NH land, whereas the maximum shrinkage rate of $0.44_{0.01}$ (0.41--0.58) million~\kmpd has occurred in the fall over the NH oceans. However, the results over the SH are different. Over land, an insignificant reduction in annual potential snowfall area is observed at 0.01 (0.04--0.02) million~\kmpd whereas, over ocean, it has increased significantly at $0.32_{0.01}$ (0.03--0.53) million~\kmpdend, despite a detected increase in average wet-bulb temperature (Fig.~\ref{Fig.5}).

\begin{table}[t]
\small
\centering
\caption{Annual and seasonal changes of hemispherical mean potential snowfall areas in million~\kmpd. The values in parentheses are percentage of changes per decade with respect to the annual or seasonal mean values for the entire length of the data. Values in bold are significant at $\alpha$ = 0.05.}
\begin{tabular}{R{1.7cm}L{1.1cm}L{1.1cm}L{1.1cm}L{1.1cm}L{1.1cm}L{1.1cm}L{1.1cm}L{1.1cm}}
\toprule
 Seasons & \multicolumn{2}{c}{ERA-Interim} & \multicolumn{2}{c}{JRA-55} & \multicolumn{2}{c}{NCEP-DOE R-2}&
\multicolumn{2}{c}{Ensemble}\tabularnewline

& Land & Ocean & Land & Ocean & Land & Ocean & Land & Ocean\\
\midrule
\multicolumn{1}{l}{\bf Northern} \\
\multirow{2}{*} {Annual}&$\bf -0.53$ &$\bf-0.28$&$\bf-0.56$& $\bf-0.39$ &$\bf-0.44$&$\bf-0.35$&$\bf-0.52$&$\bf-0.34$\\[-1pt]
& $\bf (-1.81)$ & $\bf(-1.61)$&$\bf(-1.89)$& $\bf(-2.35)$& $\bf(-1.43)$& $\bf(-2.04)$&$\bf(-1.78)$&$\bf(-1.99)$\\[3pt]

\multirow{2}{*}{Winter}& $\bf-0.31$ &$\bf-0.36$&$\bf-0.34$& $\bf-0.48$ &$-0.14$&$\bf-0.43$&$\bf-0.30$&$\bf-0.42$
\\[-1pt]
 & $\bf(-0.59)$ & $\bf(-1.41)$&$\bf(-0.64)$& $\bf(-1.93)$& $(-0.26)$& $\bf(-1.75)$ & $\bf(-0.56)$&$\bf(-1.67)$\\[3pt]
  \multirow{2}{*}{Spring} & $\bf-0.79$ &$-0.04$&$\bf-0.86$& $\bf-0.21$ &$\bf-0.58$&$-0.04$&$\bf-0.81$&$-0.10$
\\[-1pt]
 & $\bf(-2.49)$ & $(-0.19)$&$\bf(-2.65)$& $\bf(-1.03)$& $\bf(-1.67)$& $(-0.21)$&$\bf(-2.52)$&$(-0.49)$\\[3pt]
\multirow{2}{*}{Summer} & $\bf-0.30$ &$\bf-0.23$&$\bf-0.34$& $\bf-0.48$ &$\bf-0.49$&$\bf-0.35$&$\bf-0.32$&$\bf-0.34$
\\[-1pt]
 & $\bf(-7.92)$ & $\bf(-3.10)$&$\bf(-8.44)$& $\bf(-6.42)$& $\bf(-8.65)$& $\bf(-3.56)$&$\bf(-8.09)$&$\bf(-4.36)$\\[3pt]
	\multirow{2}{*}{Fall} & $\bf-0.71$ &$\bf-0.41$&$\bf-0.78$& $\bf-0.46$ &$\bf-0.37$&$\bf-0.58$&$\bf-0.70$&$\bf-0.44$
\\[-1pt]
 & $\bf(-2.49)$ & $\bf(-2.81)$&$\bf(-2.68)$& $\bf(-3.30)$& $\bf(-1.29)$& $\bf(-4.04)$&$\bf(-2.44)$&$\bf(-3.10)$\\[2pt]

 \multicolumn{1}{l}{\bf Southern} \\[-1pt]
 \multirow{2}{*}{Annual} & $-0.01$ &$\bf +0.53$&$\bf-0.02$& $+0.16$ &$\bf+0.04$&$+0.03$ &$-0.01$&$\bf +0.32$
\\[-1pt]
 & $(-0.07)$ & $\bf(+1.61)$&$\bf(-0.18)$& $(+0.50)$& $\bf(+0.33)$& $(+0.11)$&$(-0.06)$&$\bf(+1.03)$\\[3pt]
  \multirow{2}{*}{Winter}& $-0.01$ &$\bf+ 0.38$&$-0.02$& $+0.07$ &$\bf+ 0.05$&$\bf-0.28$&$-0.01$&$+0.16$
  \\[-1pt]
  & $(-0.11)$ & $\bf(+0.99)$&$(-0.19)$& $(+0.18)$& $\bf(+0.37)$& $\bf(-0.84)$&$(-0.10)$&$(+0.44)$\\[3pt]
 
 \multirow{2}{*}{Spring}& $-0.01$ &$\bf+0.34$&$\bf-0.02$& $+0.17$ &$\bf+0.03$&$-0.06$&$-0.01$&$\bf+0.26$
\\[-1pt]
 & $(-0.10)$ & $\bf(+0.91)$&$\bf(-0.14)$& $(+0.48)$& $\bf(+0.22)$& $(-0.19)$&$(-0.08)$&$\bf(+0.72)$\\[3pt]
 \multirow{2}{*}{Summer} & $\bf-0.01$ &$\bf+0.70$&$\bf-0.01$& $\bf+0.31$ &$\bf+0.01$&$\bf+0.57$&$-0.00$&$\bf+0.55$
\\[-1pt]
 & $\bf(-0.04)$ & $\bf(+2.66)$&$\bf(-0.13)$& $\bf(+1.23)$& $\bf(+0.12)$& $\bf(+2.53)$&$(-0.01)$&$\bf(+2.18)$\\[3pt]
 \multirow{2}{*}{Fall} & $-0.02$ &$\bf+0.58$&$\bf-0.02$& $+0.14$ &$\bf+0.06$&$-0.07$&$-0.01$&$\bf+0.31$
\\[-1pt]
& $(-0.14)$ & $\bf(+1.95)$&$\bf(-0.21)$& $(+0.49)$& $\bf(+0.48)$& $(-0.28)$&$(-0.08)$&$\bf(+1.08)$\\
\bottomrule

\end{tabular}

\label{Tab:2}
\end{table}
 
The contrasting nature of hemispherical changes in potential snowfall areas and average wet-bulb temperature exists due to the spatial heterogeneity of the temperature trend over the SH. Fig.~\ref{Fig.4}b shows that parts of the Southern Pacific, Atlantic and Indian oceans are warming. However, these areas are within the subtropics with no effect on the potential snowfall areas as the temperature is well above the snowfall threshold throughout the year. In fact, the major contribution to the potential snowfall areas in the SH comes from the Southern Oceans in the temperate and arctic climate zones that envelops Antarctica. Furthermore, seasonal changes are insignificant over the SH land (Table~\ref{Tab:2}) whereas over the oceans, all seasonal changes are significant -- except winter. The highest shrinkage rate occur during southern summer at $0.55_{0.01}$ (0.31--0.70) million~\kmpdend.

Mean annual potential snowfall areas are decreasing over all studied K$\ddot{\textrm{o}}$ppen-Geiger climate classes and the four mountain regions. Among the three climate classes, the highest decrease is observed over the NH cold climate regime at $0.33_{0.05}$\,(0.25--0.34) million~\kmpd followed by polar at $0.06_{0.05}$\,(0.06-0.07) and Arid cold class at $0.05_{0.01}$ (0.03--0.06) million~\kmpdend. On average, we have lost 0.02 million~\kmpd of snowfall areas over the four mountainous regions. The largest decrease is observed over the HMA at $0.05_{0.01}$ (0.04--0.06) million~\kmpd followed by the Rockies $0.01_{0.05}$ (0.00--0.02) and Andes $0.01_{0.05}$ (-0.03--0.02) million~\kmpdend. It is important to note that, the highest percentage reduction in the long-term annual mean is observed over the Alps at $3.64_{0.01}$ (-7.54--4.84)~\% pd followed by the Rockies at $2.81_{0.05}$ (1.91--3.27) and HMA at $1.85_{0.05}$ (1.53--2.18)~\% pd. 

\begin{figure}[H]
\centering\includegraphics[width=1\textwidth]{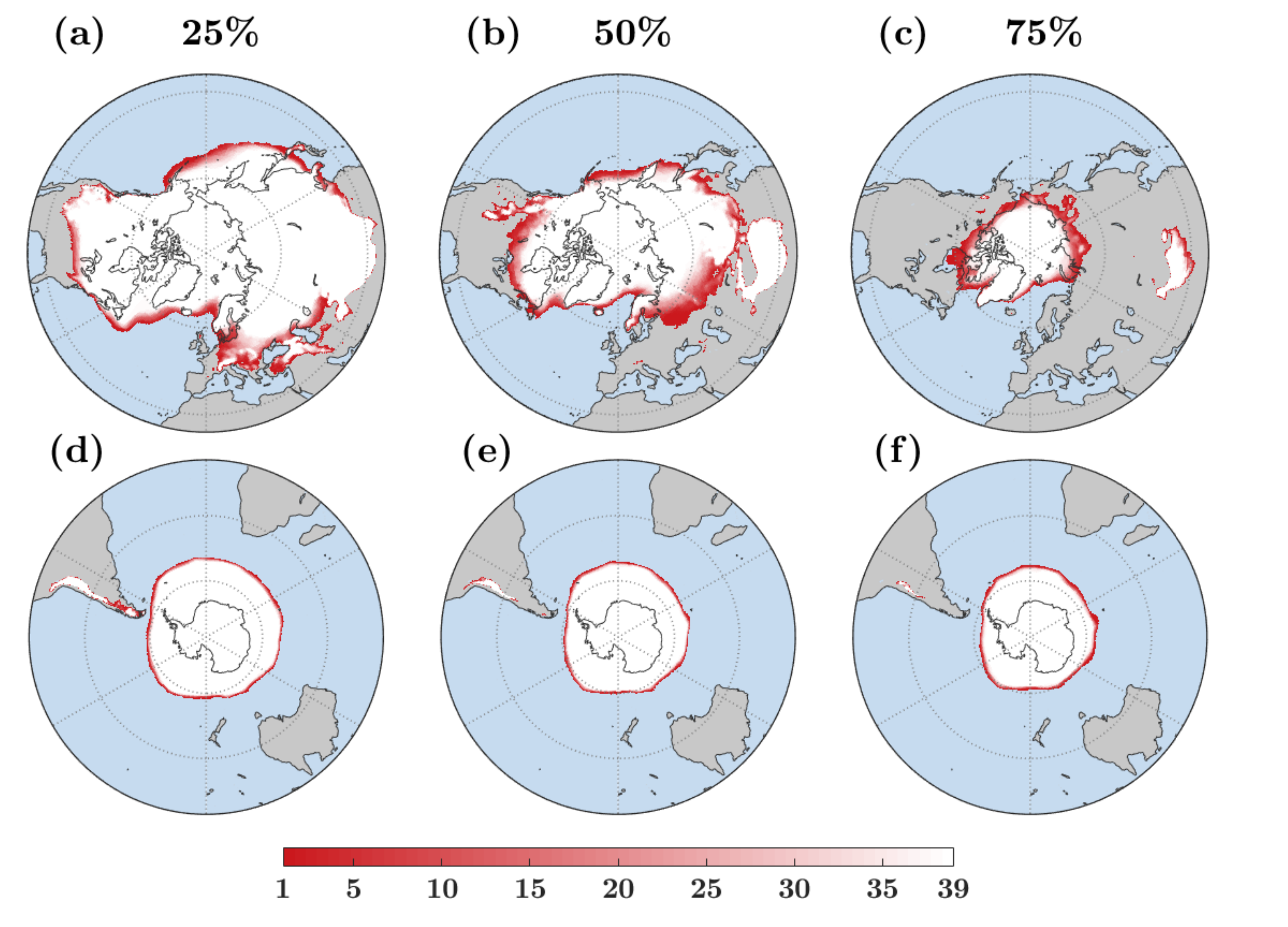}
\caption{Hemispherical frequency maps showing areas during the past four decades over the NH (a--c) and SH (d--f) -- for potential snowfall occurrences exceedance probabilities of 25\%, 50\%, and 75\%.}
\label{Fig.7}
\end{figure}

To understand where these changes occurred, we overlaid all potential snowfall areas on a daily scale throughout each calendar year using the ensemble mean wet-bulb temperatures. Then, a binary mask of the potential snowfall occurrence at a pixel-level is produced to delineate the areas that are likely to receive snowfall at least 25, 50 and 75\% of the time in a year. The frequency values obtained after overlaying the 39 years of data at these exceedance probabilities are shown in Fig.~\ref{Fig.7}~a-f at hemispheric scales. As is evident, these frequency values are likely to decrease over the areas that have potentially experienced reduced snowfall occurrence at different exceedance levels. The exceedance probabilities show significant spatial changes over the NH while they remain largely unchanged over the SH. 

Focusing on the NH terrestrial changes, among those regions where at least 25\% of the time the snowfall occurrence is likely, Eastern and Southeastern Europe, Middle East and some regions in south of the Central Asia have been experiencing significant shrinkage. From west to east, the changes extend from lowlands in Poland to the Baltic Sea, Southwest Russia, southern Kazakhstan and the Aral Sea. Over Southeastern Europe, Serbia, Bulgaria and Romania have been experiencing a shrinkage of snowfall area. In the Middle East, the changes are detected over the central west of Turkey and foothills of the Alboz and Zagros mountain ranges in the Iranian Plateau. The areas that are likely to receive snowfall more than 50\% of the time, are shrinking mostly over the North America's Rocky Mountains, Canada's boreal forests, northwest Russia and southwest Scandinavia, especially over Finland and Sweden. The shrinkage areas, with at least 75\% snowfall occurrence, are over the southern Himalayan range of HMA, eastern parts of the Tibetan Plateau and Northern Siberia. Over North America, much of the shrinkage is observed over the Brooks mountain range in Alaska and Canadian barren grounds. 

\subsection{Changes in Transition Latitudes}

The ensemble mean wet-bulb temperature enables us to understand the underlying temporal changes in the potential snowfall areas, which showed more marked changes over the NH than the SH. This is why we confine our considerations in this section only to the NH for characterizing the changes on the boundary of the potential snowfall areas. As previously noted, the motivation is to understand where and to what extent the boundaries of potential snowfall areas are moving. To that end, the snowfall to rainfall transition latitudes, representing the boundary of the potential snowfall areas, with decadal exceedance probability of 25, 50 and 75\%, are quantified. For example, the latitudes with 25\% exceedance probability represent the boundary of the areas with at least 25\% days with potential snowfall occurrences over a decade. This decadal representation is used to capture the long-term trends over a window of time and cancel out the short-term inter-annual variability. The annual zonal mean values of the transition latitudes and their poleward retraction rates are computed over $15^{\circ}$ longitude intervals. 

\begin{figure}[H]
\centering\includegraphics[width=1\textwidth]{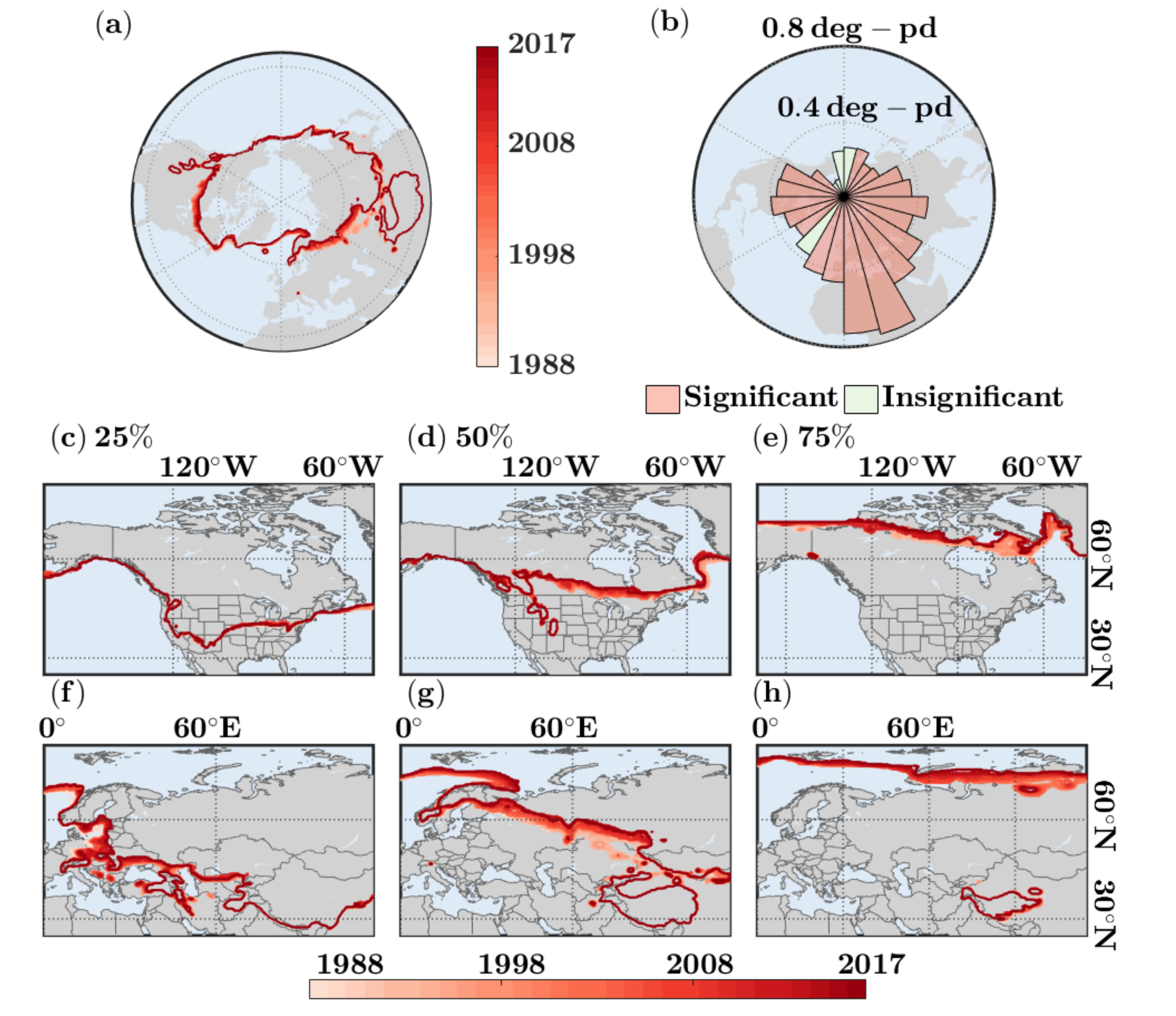}

\caption{Snowfall to rainfall transition latitudes as the boundary of potential snowfall areas with exceedance probability of 50\% over a sliding time window of 10 years (a), the annual retraction rate of the zonal mean values in degree per decade (deg-pd) (b), and the zoomed areas for different exceedance probabilities over North America (c-e) and Eurasia (f-h).}
\label{Fig.8}
\end{figure}

The transition latitudes with exceedance probability of 50\% for a moving time window of 10 years are shown in Fig.~\ref{Fig.8}a. Each line depicts the position of the boundary with lighter colors representing earlier time periods. The changes are more noticeable over land than over oceans, particularly over North America and Eurasia. Fig.~\ref{Fig.8}(c-e) and (f-h) provide a zoomed view over the areas of significant changes for decadal exceedance probabilities. 

Over North America, 25\% transition latitudes remain fairly stable with a little fluctuation in the Midwestern and Eastern United States (Fig.~\ref{Fig.8}c); however, the retraction is fairly high over the Europe and Eastern Asia (Fig.~\ref{Fig.8}f). At the 50\% level, the retraction is noticeable over the Canadian provinces of Manitoba, Ontario and Quebec (Fig.~\ref{Fig.8}d). Isolated islands of changes are formed over the Rockies and HMA (Fig.~\ref{Fig.8}d,g). Because of the steepness of the mountainous areas, areal extent changes are small and change rates should be studied as a function of elevation. Over Europe and Central Asia, a significant retraction rate is observed over the Norwegian Sea and Eastern Russia and Northern Kazakhstan (Fig.~\ref{Fig.8}g). For the 75\% level, transition latitudes are markedly retracting over Northern Quebec in Canada (Fig.~\ref{Fig.8}e), Northern Russia and over western Tibetan Plateau (Fig.~\ref{Fig.8}h). Fig.~\ref{Fig.8}(b) shows retraction rates greater than 0.7~deg-pd over Europe within 0--30 $^{\circ}$E. Also, parts of Central Asia with 75--90$^{\circ}$E, comprising the HMA, are experiencing significant retraction rate of 0.45 deg-pd.

\subsection{Snowfall to Precipitation Ratio}

\begin{figure}[t]
\centering\includegraphics[width=1\textwidth]{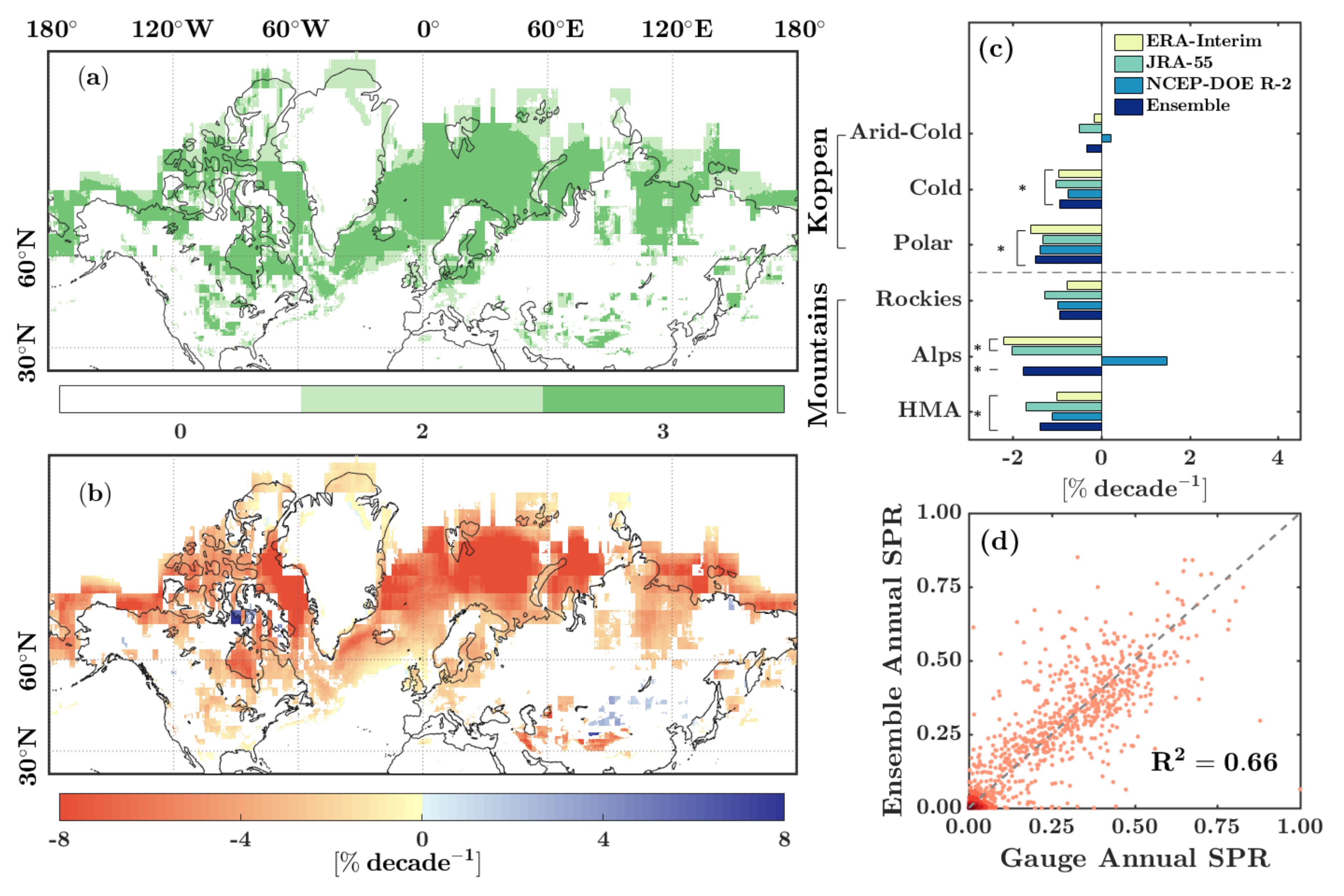}
\caption{Spatial distribution of the number of significant trends in agreement for Snowfall to Precipitation Ratio (SPR) between the three reanalysis products combined with the GPCP (a), spatial distribution of the trend in ensemble mean values of SPR in percentage per decade (b), trend in SPR for the K$\ddot{\textrm{o}}$ppen-Geiger climate classes and the mountain regions (c) and validation of the annual ensemble values of SPR with 859 station years from the GSOD gauge data (d).}
\label{Fig.9}
\end{figure}

Snowfall to Precipitation Ratio (SPR) was computed by combining cumulative precipitation information from GPCP with ancillary information of wet-bulb temperatures mapped onto the the ERA-Interim grids (0.125$^{\circ}$), using the nearest neighbour interpolation. Binary masks are produced to identify the areas where the majority of the reanalysis products and GPCP data agree that there is a decreasing or increasing trend in SPR  (Fig.~\ref{Fig.9}a). Then, the trend in annual SPR is quantified by using the ensemble mean wet-bulb temperature and GPCP data (Fig.~\ref{Fig.9}b). 

Focusing on the NH changes, it is evident that most parts of the oceans are experiencing a significant decrease in SPR around 8\% pd. However, only a few terrestrial regions exhibit significant trends around 4\% pd. Over North America, the SPR is decreasing over the Great Plains in the Unites States and  Manitoba, Ontario and Quebec provinces around the Hudson Bay in Canada. In Europe, a significant decreasing trend is detected over parts of Finland, Sweden, Poland and Germany on the coastal region of Baltic Sea as well as over the United Kingdom. Over Asia, Central Iran, Western Turkmenistan, Central Tibetan Plateau and Siberian Plateau have been experiencing a decrease in SPR in recent years, while parts of the Mongolian Plateau are experiencing a significant increasing trend of around 4\%. We need to emphasize that in general the quality of overland satellite precipitation is lower than the over ocean counterpart \citep{kubota2009verification}. This uncertainty could be one of the main reasons that the detected trends are more coherent over oceans. 

Among the three climate classes in NH, the largest decrease in the SPR occurs over the polar region at 1.49(1.33--1.59)~\% pd followed by the cold regime at 0.95 (0.75--1.02)~\% pd whereas no significant trend is observed over the arid-cold class (Fig.~\ref{Fig.9}c). The highest decrease in the SPR occurs over the Alps at 1.75 (1.48--2.22)~\% pd followed by HMA at 1.38 (1.01--1.70)~\% pd. No significant trend is observed over the Rockies.

To validate the results of the SPR calculation, the annual ensemble SPR values are compared with their counterpart from 859 NCDC gauge station years between 2011 and 2015 (Fig.~\ref{Fig.9}~d). Quality metrics including the coefficient of determination ($R^2$), relative bias (RBIAS), probability of detection (POD), false alarm ratio (FAR) and critical success index (CSI) are used as detailed in the Appendix B. The results show an $R^2$ of 0.66 and a relative bias of 0.94\%. Additionally, within the five-day GPCP temporal resolution, the probability of detection, the false alarm ratio and the critical success index are 0.9, 0.34, and 0.62, respectively.

\section{Discussion and Conclusions}
In this study, we inferred from three reanalysis data sets that the ensemble mean global wet-bulb temperature shows a significant increasing trend in the past four decades, except over the SH oceans. While all reanalysis products strongly agree on the trend over the NH, SH trends are more uncertain. There is a coherent warming trend in wet-bulb temperature over all the Arctic region whereas only parts of Antarctica, especially over the Filchner, Fimbulisen and Ross ice shelves are experiencing a warming trend. Among the three studied K$\ddot{\textrm{o}}$ppen-Geiger climate classes and various mountainous regions, the NH polar class and the Alps are experiencing the highest warming trend. 
 
Notable reductions in potential snowfall areas are observed over the NH land and oceans, especially over North America, Europe, and Northwest Russia. No significant changes have occurred over the SH land; however, a significant increase in potential snowfall areas is observed over the SH oceans despite an increase in the wet-bulb temperature. This counter-intuitive phenomenon exists because warming in the SH oceanic regions occurred mainly over the subtropics, which does not contribute to the snowfall dominant regime. However, the Southern Oceans that engulf Antarctica have experienced cooling trend thus increasing the hemispherical potential snowfall areas. On average, the four mountainous regions have experienced a reduction in potential snowfall area at 0.02 million~\kmpd with Alps losing the highest proportion of potential snowfall areas at 3.64\% pd.

Tracking the transition latitude that delineates the changes from snowfall to rainfall dominant regimes showed a retraction towards the North Pole in the NH at 0.45 deg-pd over Central Asia and 0.7 deg-pd over Europe. Furthermore, a significant reduction in SPR is observed over most parts of the NH oceans, the Great Plains of the United States, Canadian provinces around the Hudson Bay, Finland, Central Siberian and Tibetan highlands. The highest decrease in annual SPR is observed over the NH polar regime and the Alps. The results from our study are consistent with the results from station-based studies that found long term negative trend in SPR over Finland \citep{irannezhad2017long}, the Tibetan Plateau \citep{Wang2016Decrease2013}, and the contiguous United States \citep{Feng2007ChangesStates,kunkel2009trends}. 

It is worth nothing that the uncertainty of our analysis is limited to the accuracy of the data used and techniques explained. The resolution of the reanalysis data is still very coarse and the existing models and satellite data often suffer from a large degree of uncertainty over mountainous regions. Thus, the results reported over the mountains should be interpreted with caution and should be updated when new higher-resolution reanalysis data with more sophisticated parameterization of topographic feature become available. In addition, reanalysis products are highly inhomogeneous since the quality, quantity and character of assimilated data changes over time. This can introduce artificial trends that can affect our analyses \citep{long2017climatology}. Using different reanalysis products might not eliminate those trend due to the assimilation of similar data sets. Future studies could investigate snowfall changes in more homogeneous reanalyses data sets such as the ECMWF's 20th century reanalysis \citep{poli2016era} or the NOAA-CIRES's 20th Century Reanalysis \citep{compo2011twentieth} products.

Similar problem exists for the analyzed Pentad GPCP products since historical satellite data suffer from a large degree of uncertainty in retrievals of orographic precipitation. Future studies can also focus on the usage of other ancillary data such as vertical temperature lapse rate for characterizing precipitation phase to improve the inference. Moreover, finding ways to refine our inference using snowfall data from recent spaceborne active radars on board the GPM and CloudSat satellites, could be another future line of research.

{\it Acknowledgments}. The first and second authors acknowledge support from the National Aeronautics and Space Administration (NASA) Precipitation Measurement Project (NNX16AO56G), New (Early Career) Investigator Program award
(NIP, 80NSSC18K0742) and the grant from the Terrestrial Hydrology Program (THP, 80NSSC18K152). The first author also acknowledges the support provided by Sommerfield Graduate Fellowship at University of Minnesota-Twin Cities during his first year of study. NCAR is sponsored by the National Science Foundation.

\centerline{\Large Appendix A}
\centerline{\large \bf{Wet-bulb Temperature Computation}}
\noindent Here, the wet-bulb temperature is calculated from equation \ref{Eqn:Wb}  \citep{stull2016practical} using an iterative Newton-Raphson method:

\begin{equation}
     T_w=T_a-\frac{L_v}{C_p}\cdot\mathcal{E}\cdot A\cdot \bigg{(}\frac{1}{P \cdot \exp({\frac{B}{T_w}})-A}-\frac{RH}{P\cdot \exp({\frac{B}{T_a}})-A}\bigg{)}
     \label{Eqn:Wb}
\end{equation}

where, $T_w$ is the 2 m wet-bulb temperature in Kelvin (K); $T_a$ denotes 2 m air temperature in K; $L_v$ is the latent heat of vaporization in J/kg; $C_p$ refers to the specific heat constant for dry air in KJ/(Kg$\cdot$K); $\mathcal{E}$ is the ratio of dry gas constant to water vapor gas constant; $P$ denotes station pressure in millibar (mb); RH is the relative humidity; and $A=2.53\times 10^9$ mb and $B=5420$ K are empirical constants.

Due to the lack of information on RH within ERA-Interim, it was computed from dew-point temperature and air temperature using Eq.~\ref{eq:RH} provided by \citet{ECMWF2015IFSEngland}.

\begin{equation}
\label{eq:RH}
\begin{split}
RH= \exp⁡ \bigg[17.502 \bigg(\frac{T_d-273.16}{T_d-32.19}-\frac{T_a-273.16}{T_a-32.19}\bigg)\bigg]
\end{split}
\end{equation}

where, $T_d$ is the dew-point temperature in K.

\centerline{\Large Appendix B}
\centerline{\large \bf{Quality Metrics}}
\begin{table}[H]
\centering
\caption{Statistical measures for performance evaluation of ML estimate}
\begin{tabular}{lcc}
\toprule
Indices & Formula & Range\\
\midrule
Coefficient of Determination ($R^2$)  & $\displaystyle \bigg\{ \frac{1}{N_s-1} \sum_{i=1}^{N_s}\bigg(\frac{\hat{x}_{ML}(i)-\mu_{ML}}{\sigma_{ML}}\bigg)\bigg(\frac{x_G(i)-\mu_G}{\sigma_G}\bigg)\bigg\}^2$& 0 -- 1 \\
\\
Relative Bias (RBIAS) & $\displaystyle  \frac{\sum_{i=1}^{N_s}(\hat{x}_{ML}(i)-x_G(i))}{\sum_{i=1}^{N_s}x_G(i)} \cdot 100$ & -$\infty$ -- $\infty$\\
\\
Probability of Detection (POD) & $\displaystyle\frac{n_H}{n_H+n_M}$ & 0--1\\
\\
 False Alarm Ratio (FAR) &$\displaystyle\frac{n_F}{n_H+n_F}$& 0--1\\
 \\
Critical Success Index (CSI)&$\displaystyle\frac{n_H}{n_H+n_M+n_F}$& 0--1 \\
\bottomrule

\end{tabular}

\label{Tab:Appendix}
\end{table}
\noindent The used quality metrics in this study are shown in Table \ref{Tab:Appendix} as follows:

\noindent where, $\hat{x}_{ML}(i)$ is the maximum likelihood estimate for $i^{th}$ station day (or year) with standard deviation $\sigma_{ML}$, $x_G(i)$ is the corresponding gauge observation with standard deviation $\sigma_G$; $N_s$ is the total number of station days (or years), $\mu_{ML}$ and $\mu_G$ are the mean of ML estimates and gauge observations respectively. In computation of false alarm ratio and critical success index, $n_H$ is the number of hits, $n_M$ is the number of misses and $n_F$ is the number of false alarms.

\bibliographystyle{ametsoc2014}
\bibliography{shared.bib}

\end{document}